\documentclass{PoS}
\usepackage{graphicx}
\usepackage{subcaption}
\usepackage[utf8]{inputenc}
\usepackage[export]{adjustbox}
\usepackage{wrapfig}
\usepackage{amsmath}

\title{Studies of the Time Structure of Extended Air Showers for Direction Reconstruction with the HAWC Outrigger Array}

\ShortTitle{Curvature Study}

\author{\speaker{Dezhi Huang}\\
        Michigan Technological University, Houghton, Michigan, USA\\
        E-mail: \email{dezhih@mtu.edu}}

\author{For the HAWC Collaboration\\
     For collaboration list and acknowledgement: PoS(ICRC2019)1177 
     \href{https://www.hawc-observatory.org/collaboration/icrc2019.php}{https://www.hawc-observatory.org/collaboration/icrc2019.php}}
\abstract{The High Altitude Water Cherenkov (HAWC) gamma-ray observatory is a ground-based air shower array designed to detect Cherenkov light produced in water by secondary particles from atmospheric air showers. In order to improve the sensitivity at the highest energies, especially for the shower cores falling outside the main array, 345 smaller Water Cherenkov Detectors (WCDs) were installed around the main array, the outrigger array. This extension increased the instrumented area of HAWC by a factor of four. With the increased size of the array, and the ability to detect shower particles further away from the core, understanding of the time structure of the shower front is crucial for accurate direction reconstruction and mandates proper modeling. In this contribution, we present a model of the shower front as expected to be observed by the outriggers obtained from Monte-Carlo simulations. Applying this model to shower reconstruction, the improvements on air shower parameter are studied.}

\FullConference{36th International Cosmic Ray Conference -ICRC2019-\\
		July 24th - August 1st, 2019\\
		Madison, WI, U.S.A.}

\begin{document}

\section{Introduction}
\begin{figure}
    \centering
    \includegraphics[scale=0.4]{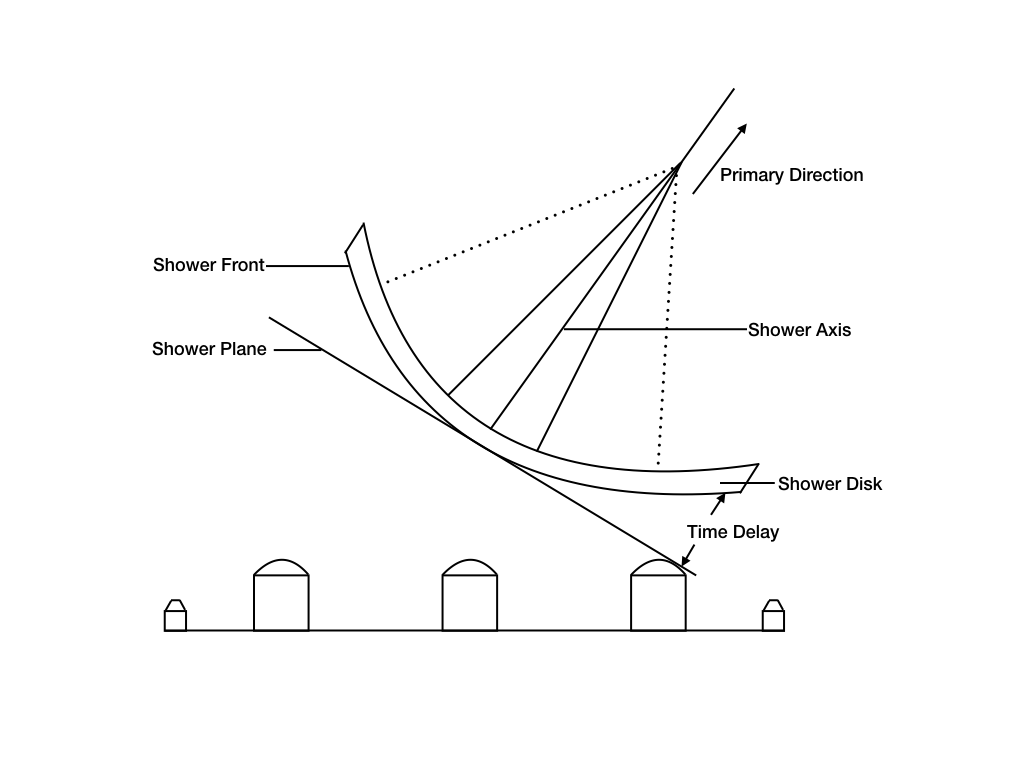}
    \caption{The curved air shower front}
    \label{fig:curvature}
\end{figure}

When a primary high energy gamma-ray event enters into the atmosphere, it may interact with matter and produce an electron-positron pair. The electron and positron will be further deflected in the electric fields of atomic nuclei losing energy by emitting gamma-ray photons. Eventually, a primary high energy gamma-ray photon will develop into a cascade of secondary particles that can reach ground level. Typically, the arrival time distribution of air showers has a cone like structure with a curved shower front as illustrated in Figure~\ref{fig:curvature}. The particle density decreases with distance to the shower axis and at larger distances particles usually are subject to more scattering interactions~\cite{grieder2010extensive}. In the angular reconstruction procedure we fit a shower plane by minimizing a $\chi^2$ function. The shower plane propagates with the speed of light along the shower axis and the shower direction is perpendicular to the particle plane. 
\par
HAWC is a ground-based air shower detector designed to observe Cherenkov light produced in water by secondary particles from air showers. With the newly installed small WCDs around the main array~\cite{outriggers_vincent}, HAWC has the ability to capture air showers that fall outside of the main array more precisely.  In this proceeding, we investigate the shower structure detected by the HAWC main array and the outrigger WCDs with simulated data. Precise studies of shower structure will eventually benefit the angular reconstruction for the HAWC observatory. 
\section{Simulation of Gamma-ray Air Showers}
\begin{figure}
    \centering
    \includegraphics[scale=0.5]{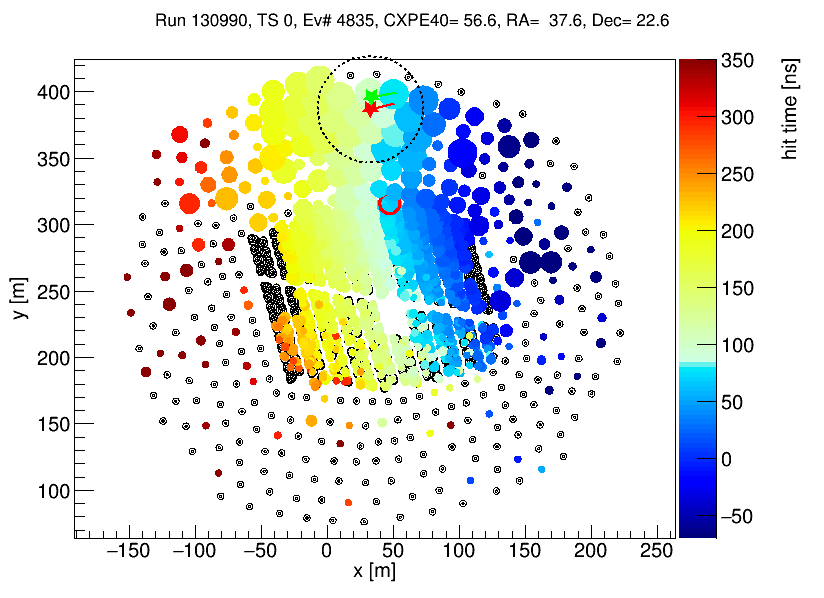}
    \caption{A simulated gamma-ray event hitting both the main array and the outrigger WCDs. The green star mark is the true core location and the red star mark is the reconstructed core. The dished circle centered on the red star has a radius of 40 meters. The small red circle near the center of detector is the maximum charge falling outside the dashed circle. The color scale represents the time air shower particles trigger a PMT in the array.}
    \label{Simulated gamma-ray event}
\end{figure}

The gamma-ray air showers are simulated with CORSIKA~(v7.4000)~\cite{heck1998corsika} down to a horizontal plane 50 meters above the HAWC tanks. Then the particles from the CORSIKA showers are propagated from this injection plane onto the main array and the outrigger WCDs by GEANT 4~\cite{agostinelli2003geant4}. GEANT 4 is also used to track the Cherenkov light to the top of the photomultiplier tubes~(PMTs). The response of the PMTs and data acquisition system~(DAQ) are also simulated with the custom-made code. Then the reconstruction~\cite{outriggers_vikas} is performed as for real air shower data. All the simulations used here are for gamma-ray events. Figure~\ref{Simulated gamma-ray event} shows a simulated gamma-ray event with 141 TeV energy, the shower core falls outside the main array but is captured by the outriggers.

\section{Angular Reconstruction and Curvature Correction}
In this section, the method used to derive the shower curvature information will be introduced. In the current HAWC reconstruction procedure~\cite{abeysekara2017observation} the "center of mass" of the PMT charges is calculated first. It provides an initial guess of the shower core location. This core location is used in the angular reconstruction and a shower plane is fit by minimizing a $\chi^2$ function~\cite{atkins2003observation}. 
\begin{equation}
\chi^2=\displaystyle\sum_{n=1}^{N} \omega_n(ct_n-ct_0+ix_n+jy_n+kz_n)^2,
\end{equation}
where N is number of hits, $\omega_n$ is a weight parameter based on inverse charge measured by the PMTs contributing to air shower, $t_n$ is the arrival time of each hit after curvature correction and $t_0$ is absolute arrival time of the shower. The parameters $x_n$, $y_n$, $z_n$ denote the hit location in detector coordinates and $i$, $j$, $k$ are the components of the unit vector corresponding to the shower axis:

\begin{equation}
    \left(
    \begin{array}{c}
      i \\
      j \\
      k
    \end{array}
  \right) = \left(
    \begin{array}{c}
      cos\theta sin\phi \\
      sin\theta sin\phi \\
      cos\theta
    \end{array}
  \right) 
\end{equation}
Here, $\theta$ is the zenith angle and $\phi$ is the azimuth angle of the shower axis. The direction of the primary particle is perpendicular to this plane. 

In the current HAWC angular reconstruction procedure, there are two effects that will be corrected before the $\chi^2$ plane fit is performed. The first one is the curvature. As mentioned in the first section, the shower front has a slight conical shape centered at the shower axis~\cite{abeysekara2017observation}. The shower will become thicker when particles are further from the shower axis. This second effect is referred as "sampling": the electronic channels measure the arrival time of the first arriving photoelectron and thus, the measurement time will be close to the true shower front for high particle rates, but tends to be delayed for low particle rates. A combined curvature and sampling correction will be used to correct those two effects. These corrections are derived from a combination of simulations and real gamma-ray air shower events collected from the direction of the Crab nebula.

\begin{figure}[!htbp]

\begin{subfigure}{0.45\textwidth}
\includegraphics[width=1\linewidth, height=5.5cm]{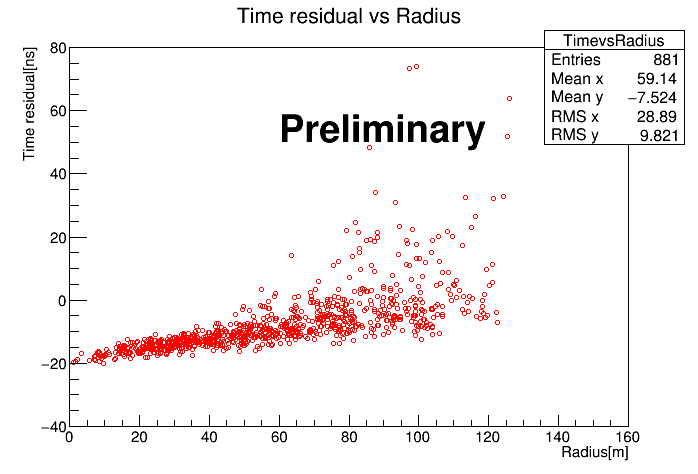} 
\caption{Main array hits}
\label{fig:subim1}
\end{subfigure}
\begin{subfigure}{0.45\textwidth}
\includegraphics[width=1\linewidth, height=5.5cm]{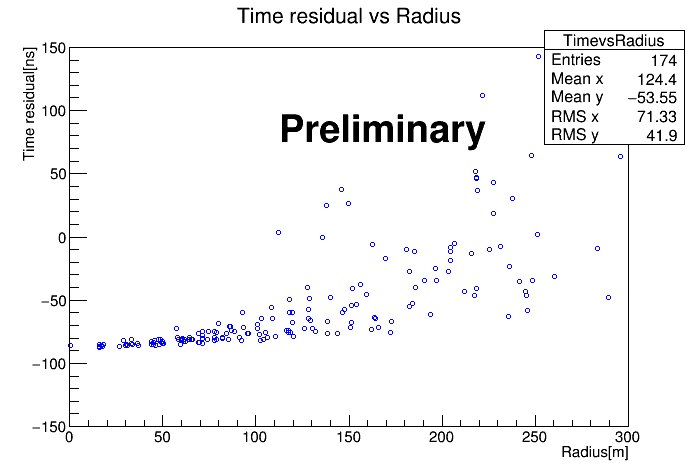}
\caption{Outrigger hits}
\label{fig:subim2}
\end{subfigure}

\caption{Time structure of the main array and the outrigger WCDs from a simulated gamma-ray event}
\label{3}
\end{figure}

\begin{figure}[!htbp]

\begin{subfigure}{0.45\textwidth}
\includegraphics[width=1\linewidth, height=5.5cm]{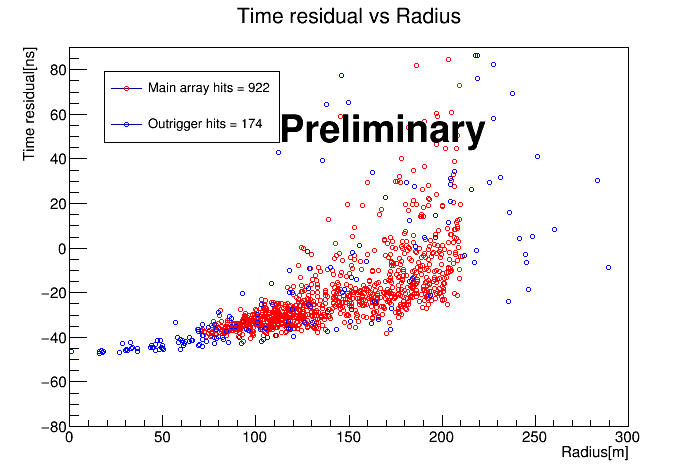} 
\caption{Before correction}
\label{fig:subim1}
\end{subfigure}
\begin{subfigure}{0.45\textwidth}
\includegraphics[width=1\linewidth, height=5.5cm]{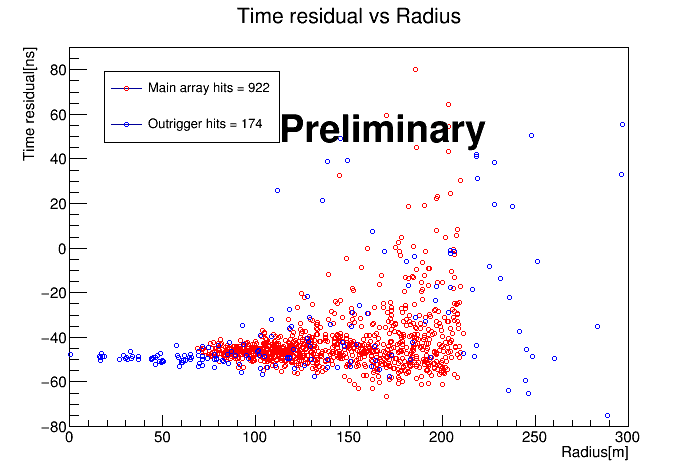}
\caption{After correction}
\label{fig:subim2}
\end{subfigure}
\caption{Time structure of the main array and the outrigger WCDs, the left plot is before and the right plot is after HAWC curvature and sampling correction}
\label{4}

\end{figure}

We can calculate the time difference between PMT hits and arrival times expected from the shower plane fit. In Figure~\ref{3}, we can see the time structure of the shower curvature of a simulated gamma-ray event. Figure~\ref{4} shows the time structure of one simulated event displayed in Figure 2 before and after applying the combined curvature and sampling correction. The curve becomes flatter. However, we can still find some late tails even after correction.

\section{Apply HAWC Curvature Correction to the Outriggers}

The sampling correction introduced in section 3 depends on the particle rate in each tank. In the current procedure, the measured charge is used as a proxy for the particle rate. Since the outrigger WCDs are much smaller than the main array WCDs, the relationship between particle rate and charge will be different. Also, in the main array Cherenkov light from particles that hit at the edge and the center of the tank will have different time delays. For the outriggers the difference will be much smaller than the main array. Even though the shower structure is the same, we could still measure a different time structure between the main array and the outriggers because of the difference in the sampling correction. 

In order to investigate the main array and the outriggers separately, we calculate the time residual for the outriggers and the main array hits. Then we choose a 40 ns time window before and after the shower plane arrival to detector. We fit a linear function after curvature and sampling correction:
\begin{equation}
    y=const+\lambda r
\end{equation}
Where $y$ is the time delay from Figure 1, and r is the distance from hit PMTs to reconstructed core. The constant, $const$ and $lambda$ are the fit parameters.

\begin{figure}[!htbp]
\begin{subfigure}{0.45\textwidth}
\includegraphics[width=1\linewidth, height=5.5cm]{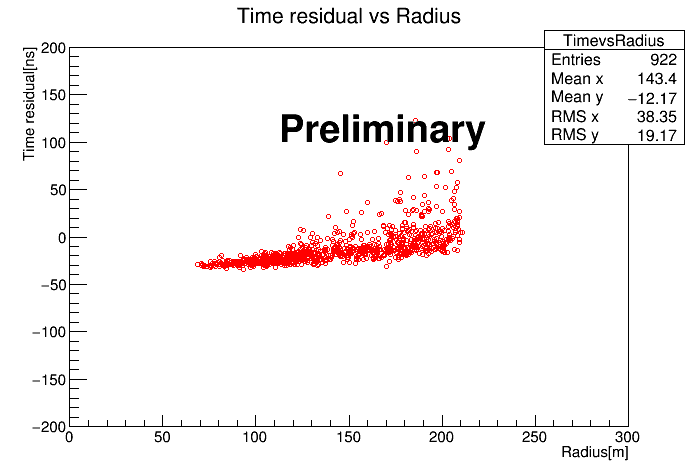} 
\caption{Main array hits before correction}
\label{fig:subim1}
\end{subfigure}
\begin{subfigure}{0.45\textwidth}
\includegraphics[width=1\linewidth, height=5.5cm]{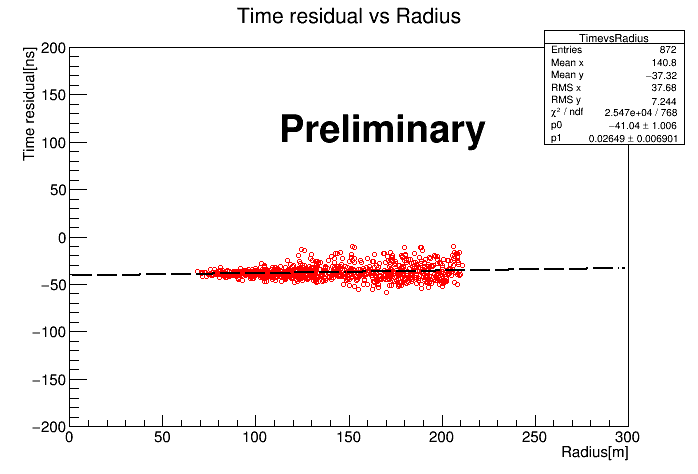}
\caption{Main array hits after corrections}

\label{fig:subim2}
\end{subfigure}

\begin{subfigure}{0.45\textwidth}
\includegraphics[width=1\linewidth, height=5.5cm]{130990_out.png} 
\caption{Outrigger hits before correction}
\label{fig:subim3}
\end{subfigure}
\begin{subfigure}{0.45\textwidth}
\includegraphics[width=1\linewidth, height=5.5cm]{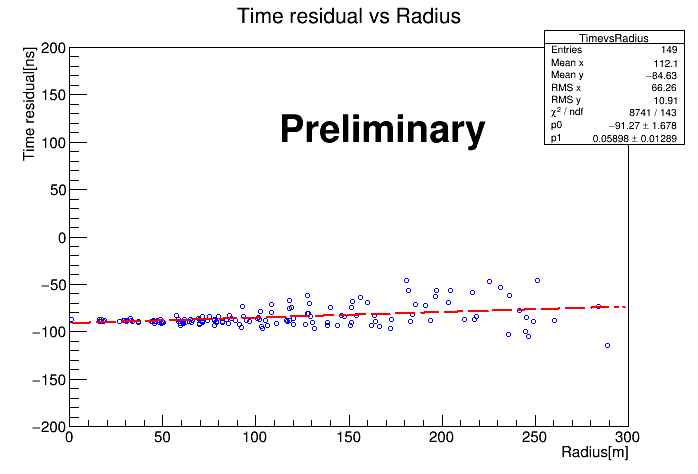}
\caption{Outrigger hits after correction}
\label{fig:subim4}
\end{subfigure}
\caption{Time structure of the outrigger hits the left plot is before correction and the right plot is after HAWC curvature correction.}
\label{5}
\end{figure}
Figure~\ref{5} shows the main array hits and the outriggers hits before and after correction for the curvature and the sampling effects. The fit is performed after applying the correction. This preliminary study shows there is a slight difference between the main array slope and the outriggers slope. Further studies are in progress. 
\section{Conclusion}
In this proceeding, we investigated the shower structure for the HAWC main array and the outrigger detectors. We applied the standard HAWC curvature and sampling correction to both the main array and the outrigger hits. With this preliminary study, we found that some structure remains after correction and the slope between the main array and the outriggers hits have a slight difference. In the future we will analyze larger samples to investigate the charge dependence and spatial dependence of the time delay for both main array and outrigger tanks. The goal is to improve HAWC's angular resolution.

\section{Acknowledgement}
Thanks for the support from the US National Science Foundation (NSF), my advisor Petra Huentemeyer and my colleagues: Chad Brisbois, Henrike Fleischhack, Binita Hona at Michigan Technological University.

\end{document}